\documentclass[aps,preprint,showpacs]{revtex4}
\usepackage{amsmath}
\usepackage{graphics}

\begin{document}
\bibliographystyle{/usr/share/texmf/tex/latex/revtex4/prsty}

\title{\hfill{\small {\bf MKPH-T-04-09}}\\
Ab initio calculation of $^7$Li photodisintegration}

\author{Sonia Bacca$^{1,2}$, Hartmuth Arenh\"{o}vel$^2$, 
Nir Barnea$^3$, Winfried Leidemann$^{1}$ and Giuseppina Orlandini$^{1}$}

\affiliation{$^1$Dipartimento di Fisica, Universit\`a di Trento 
and INFN (Gruppo Collegato di Trento),\\
via Sommarive 14, I-38050 Povo, Italy\\ 
$^2$Institut f\"ur Kernphysik, Johannes Gutenberg-Universit\"at
Mainz,\\ 
Johann-Joachim-Becher-Weg 45, D-55099 Mainz, Germany\\
$^3$Racah Institute of Physics, Hebrew University, 91904, 
Jerusalem, Israel}

\date{\today}

\begin{abstract}
The $^7$Li total photoabsorption cross section is calculated microscopically.
As nucleon-nucleon interaction the semi-realistic central AV4' potential 
with S- and P-wave forces is taken. The interaction of the final 7-nucleon 
system is fully taken into account via the Lorentz Integral Transform (LIT) 
method. For the calculation of the LIT we use expansions in hyperspherical 
harmonics (HH) in conjunction with the HH effective interaction (EIHH) 
approach. The convergence of the LIT expansion is discussed in detail. 
The calculated cross section agrees quite well with the available 
experimental data, which cover an energy range from threshold up to 100 MeV.
\end{abstract}
\bigskip
\maketitle

\vfill\eject

The electromagnetic response of an $A$-body nucleus is a basic property in 
nuclear physics. It contains important information about the dynamical 
structure of the system. For this reason microscopic calculations are 
needed in order to investigate the details of the reaction mechanism 
and the underlying dynamics.
Traditionally such studies could only be made for few-body systems with 
$A \leq 3$. The  explicit calculation of final 
state wave functions constitutes the main limitation of standard approaches. 
On the other hand these difficulties 
can be avoided in the Lorentz Integral Transform (LIT)  method \cite{LIT}, 
where only a bound-state-like problem has to be solved although the 
full final state interaction (FSI) is rigorously taken into account.
First applications of the method were made for electromagnetic responses of
$^4$He \cite{4He}.
Enormous progress has also been achieved in the calculation of bound-state wave 
functions of systems with $A>4$ (No Core Shell Model \cite{NCSM} and 
Effective Interaction Hyperspherical Harmonics (EIHH) approach~\cite{EIHH}). 
The novel methods of LIT and EIHH combined with the help of 
modern computational resources enabled us to carry out calculations of 
electromagnetic cross sections of nuclei with $A=6$ \cite{Sonia02,Sonia04}.
Therefore, systems with $4<A<12$ are now considered an interesting 
playground for 
establishing a bridge between few- and many-body physics and especially for 
testing many-body approximations.  

In this paper we present the first microscopic calculation
of a photoabsorption reaction on a nucleus with A=7, namely $^7$Li.
In this calculation we use as nucleon-nucleon interaction the 
Argonne potential AV4' \cite{AV4'}, which is a semirealistic central 
interaction that contains S- and P-wave forces.
In recent work~\cite{Sonia04} we found that the P-wave interaction has 
a significant influence on the photodisintegration cross section of 
P-shell nuclei, like $^6$He and $^6$Li, and leads to a considerably better
agreement with experimental data than a central S-wave interaction only. Thus 
we have chosen this potential model for the 
present ab initio calculation of the $^7$Li total photoabsorption cross
section. The calculational procedure is described in detail in 
\cite{Sonia02, Sonia04}. In the following we will briefly review 
the main steps only.

The inclusive unpolarized response function $R(\omega )$ is defined by 
the total photoabsorption cross section 
\begin{equation}
\label{0}
\sigma(\omega)=4\pi^{2}\alpha\omega R(\omega)\,,
\end{equation}
where $\alpha$ denotes the fine structure constant and $\omega$  the 
photon energy. In the LIT method~\cite{LIT} one obtains $R(\omega)$ from 
the inversion of the integral transform
\begin{equation}
\mathcal{L}(\sigma_{R},\sigma_{I} )= \int d\omega \frac{R(\omega )}
{(\omega -\sigma _{R})^{2}+\sigma ^{2}_{I}} \,.
\end{equation}
In the unretarded dipole approximation one has
\begin{eqnarray}
\mathcal{L}(\sigma_{R},\sigma_{I} )
& = &
\frac{1}{2L_0+1}\,\,\frac{1}{2S_0+1}\sum_{M_0^L, M_0^S} 
\frac{1}{\sigma_I}\cr 
& \times &{\rm Im}\{  \langle 
\Psi_0 (M_0^L,M_0^S)|\hat{D}^{\dag}_z \frac{1}{H-\sigma_R-i\sigma_I} 
\hat{D}_z | \Psi_0 (M_0^L,M_0^S)\rangle   \}
, \label{2}
\end{eqnarray}
where $\hat{D}_z$ denotes the dipole operator, $H$ the nuclear 
Hamiltonian, and $L_0$ and $S_0$ are the angular momentum and spin
of the ground state $| \Psi_0 (M_0^L,M_0^S)\rangle$,
 with projections $M_0^L$ and $M_0^S$, respectively
(since we work with central forces, $L_0$ and $S_0$ are good quantum 
numbers). 
The transform $\mathcal{L}(\sigma_{R},\sigma_{I})$ is evaluated by 
inserting 
a complete set of projection operators $\sum_{C,M^L,M^S}|\Psi_C(M^L,M^S)
\rangle\langle \Psi_C(M^L,M^S)|\equiv\sum_{C,M^L,M^S}
{\hat P}_{C,M^L,M^S}$,
where $C=\{L,S,T,T^z,\pi\}$ stands for the quantum numbers 
characterizing the channels (angular momentum, spin, isospin and its 
projection, and parity, respectively), $M^L$ and $M^S$ are third 
components of angular momentum and spin. 
In the sum only the channels
allowed by the dipole selection rules need to be considered and 
since the dipole operator does not depend on spin,
we do not need to average over the initial spin projections ($M_0^S$),
neither to sum over $M^S$. Therefore one obtains 
\begin{eqnarray*}
\mathcal{L}(\sigma_{R},\sigma_{I} )&=&\frac{1}{2 L_0+1}
\frac{1}{\sigma_I}\sum_{C}
|\langle 
\Psi_C |  \hat{D}_z  | \Psi_0 \rangle |^2
{\rm Im}\{  \langle 
\Psi_C |\frac{1}{H-\sigma_R-i\sigma_I}| \Psi_C \rangle \},
\end{eqnarray*}
where  

\begin{equation}
 | \Psi_C  \rangle ={\frac{{\hat P}_{C} \hat{D}_z | 
\Psi_0 \rangle}
       {\sqrt{\langle \Psi_0|\hat{D}_z {\hat P}_
C\hat{D}_z| \Psi_0\rangle}}}.
\end{equation}
To simplify the notation, in the last two equations we have also omitted
the dependences on $M^L$ and $M_0^L$, as will be done in the following. 

In order to use the Lanczos algorithm (see Ref.~\cite{Mar02}) it is 
convenient to write
$\mathcal{L}(\sigma_{R},\sigma_{I} )=\sum_C {\cal N}^2_C F_C $, where 
${\cal N}_C$
is connected to the norm of ${\hat P}_{C} \hat{D}_z| \Psi_0 \rangle$, 
\begin{eqnarray}\label{n_C}
{\cal N}^2_C & = &
\frac{1}{2L_0+1}\sum_{M^L, M_0^L}
|\langle \Psi_C |  \hat{D}_z  | \Psi_0 \rangle |^2 =
\frac{1}{2L_0+1}\sum_{M^L, M_0^L} 
\langle 
\Psi_0|\hat{D}_z {\hat P}_C
\hat{D}_z | \Psi_0
\rangle
\end{eqnarray}
and
\begin{eqnarray} \label{7}
 F_C &=& \frac{1}{\sigma_I}{\rm Im}\{  \langle 
         \Psi_C |\frac{1}{H-\sigma_R-i\sigma_I}| \Psi_C  \rangle \}
\end{eqnarray}
is evaluated as continuous fraction via the Lanczos coefficients. 
This means that the  LIT of the total response is the sum of the 
Lorentz transforms of the individual channels 
\begin{equation}
\mathcal{L}(\sigma_R,\sigma_I)=\sum_C \mathcal{L}_C(\sigma_R,\sigma_I)\,, 
\end{equation}
where for every channel one has $\mathcal{L}_C(\sigma_R,\sigma_I)={\cal N}^2_CF_C$.

In the case of $^7$Li one has total angular momentum and parity $J^{\pi} = 
\frac{3}{2}^-$ and isospin $T_0=\frac{1}{2}$ with projection $T_0^z=
-\frac{1}{2}$. Using central forces, one has ground state orbital angular 
momentum $L_0=1$ and spin $S_0=\frac{1}{2}$. There are six different channels 
allowed by the dipole selection rules corresponding to angular momentum $L=
L_0-1, L_0, L_0+1$, spin $S=S_0$ and isospin $T=T_0, T_0+1$ with isospin 
projection conserved  $T^z=T^z_0$. In Table I we show the good quantum 
numbers of these channels.
\begin{table}
\caption{Good quantum numbers for the channels $|\Psi_C\rangle$ allowed by 
the dipole selections rules.}
\begin{ruledtabular}
\begin{tabular}{c|cccccc}
 & $|\Psi_1\rangle$ & $|\Psi_2\rangle$ & 
$|\Psi_3\rangle$ & $|\Psi_4\rangle$ & $|\Psi_5\rangle$ & $|\Psi_6\rangle$\\
\colrule
$L$ & $0$ & $0$ & $2$ & $2$ & $1$ & $1$\\
$S$ & $\frac{1}{2}$ & $\frac{1}{2}$ & $\frac{1}{2}$ & 
$\frac{1}{2}$ & $\frac{1}{2}$ & $\frac{1}{2}$\\
$T$ & $\frac{1}{2}$ & $\frac{3}{2}$ & $\frac{1}{2}$ & 
$\frac{3}{2}$ & $\frac{1}{2}$ & $\frac{3}{2}$\\
$T^z$ & $-\frac{1}{2}$ & $-\frac{1}{2}$ & 
$-\frac{1}{2}$ & $-\frac{1}{2}$ & $-\frac{1}{2}$ & $-\frac{1}{2}$\\
\end{tabular}      
\end{ruledtabular}                                                             
\end{table}

For every channel the LIT is calculated by expanding $|\Psi_{0}\rangle$ 
and $|\Psi_C\rangle$ in terms of the 7-body 
anti-symmetrized hyperspherical harmonics (HH) up to a maximum HH 
hyperangular momentum~\cite{BN97}, i.e.\ $K_{max}^0$ for the ground state 
$|\Psi_0\rangle$ and $K_{max}$ for the channels $|\Psi_C\rangle$. The 
convergence of the HH expansion is improved by using the EIHH approach. Our 
procedure consists in fixing first $K^0_{max}$ such that convergence for the 
binding energy is reached, and then we study the behavior of the LIT with 
increasing $K_{max}$. As already pointed out in \cite{Sonia04}, the rate of 
convergence for a potential that includes a central P-wave interaction, like 
the AV4', can be slower than that of a purely central S-wave force. 
Nevertheless, in the case of $^7$Li we have obtained a satisfactory 
convergence in 
terms of hyperangular momentum, both for ground state energy and LIT. For the 
ground state of $^7$Li good convergence is reached with $K^0_{max}= 9$ with a 
binding energy of 45.28 MeV. Further increase to $K^0_{max}= 11$ leads only to 
a small change of the binding energy by 0.05 MeV. 
Because of the dipole selection rule
$K=K_0\pm 1$ between states with hyperangular momenta $K$ and $K_0$, an 
expansion of the ground state up to a certain $K^0_{max}$ implies that in
$|\Psi_C\rangle$ only states with hyperangular momentum 
$K\leq K_{max}=K^0_{max}+1$ contribute to the LIT. Thus it is expected 
that for sufficiently high $K^0_{max}$ a further increase of $K_{max}$ 
beyond $K^0_{max}+1$ will not result in a significant change. In this case
a check of the convergence with respect to $K^0_{max}$ only will be 
sufficient.


In the present work, the best calculation of the LIT corresponds to a final
state with $K_{max}=10$. In this case the number of HH-basis states to be 
included becomes quite large, especially for the channels with $L_C=1$ and 2.
For example, already for channel $C=3$ and $K_{max}=10$ one has 
6348 hyperspherical states. This number has to be multiplied by the 
number of hyperradial states, about 30, to obtain the total number 
of states needed in the expansion. 
Therefore, it is desirable to discard those HH states which give only  
negligible  contributions to the LIT. To this end we have studied the 
importance of the HH states according to their spatial symmetry and found that 
quite a few of them can be  safely neglected (see Table II). We have checked 
this approximation by performing calculations with the complete set of states 
for lower values of $K_{max}$ (6 and 8) and compared the results with those 
using a truncated set. Whenever the differences between results with the 
reduced and the full basis were negligible (below $0.5\%$) we concluded that 
the omitted states were negligible also for higher $K_{max}$. 
In this way we accomplished for $K_{max}=10$ a sizable reduction from 
$N=190\,440$ to $N=111\,900$ basis functions.
\begin{table}
\caption{Reduction of basis states for channel $C=3$ for $K_{max}=10$. 
Each symmetry is represented by a Young tableau. The total number of states
for each symmetry is listed as $N^{sym}$, while $N^{sym}_{used}$ lists 
the actual number of states in the calculation.}
\begin{ruledtabular}
\begin{tabular}{c|ccccccccccc}
Symmetry & [1111111] & [211111] & [22111] & [31111] & [2221] & [3211] 
& [4111] & [322] & [331] & [421] & [43]\\
\colrule
$N^{sym}$ & 5 & 69 & 240 & 314 & 315 & 977 & 693 & 698 & 780 & 1515 & 742\\
$N^{sym}_{used}$ & 0 & 0 & 0 & 0 & 0 & 0 & 693 & 0 & 780 & 1515 & 742\\
\end{tabular}      
\end{ruledtabular}
\end{table}
In an analogous way we carried out the calculations for all the other 
channels. The estimated error introduced by these truncations is of 
the order of $0.5\%$. 

We are also able to check the error introduced by the symmetry 
truncation in a second 
way. In fact a good check of the quality of the calculation is obtained by 
considering the sum over the norms ${\cal N}_C^2$ defined in (\ref{n_C}). Using 
completeness one finds
\begin{eqnarray}
\sum_C {\cal N}^2_c=\frac{1}{2L_0+1}\sum_{C, M^L, M_0^L}|\langle 
\Psi_C |  \hat{D}_z  | \Psi_0 \rangle |^2=\frac{1}{2L_0+1} \sum_{M_0^L}\langle 
\Psi_0 |  \hat{D}^{\dag}_z  \hat{D}_z  | \Psi_0 \rangle \,,
\label{8}
\end{eqnarray}
where the last expression is nothing else than the mean expectation 
value of the operator 
$\hat{D}^{\dag}_z  \hat{D}_z$ in the ground state, that can be easily 
calculated (see Ref.~\cite{Sonia02}). 
With respect to Eq.~(\ref{8}) we obtained $1.877~[{\rm fm}^2]$ 
for the ground state expectation value with $K^0_{max}=9$, 
while using a symmetry truncated expansion for $|\Psi_C\rangle$ 
up to $K_{max}=10$ 
we get $1.871~[{\rm fm}^2]$. The small difference of $0.3\%$ reflects 
the small error introduced by the symmetry truncation.  

\begin{figure}[ht]
{\resizebox*{8cm}{12cm}{
\includegraphics{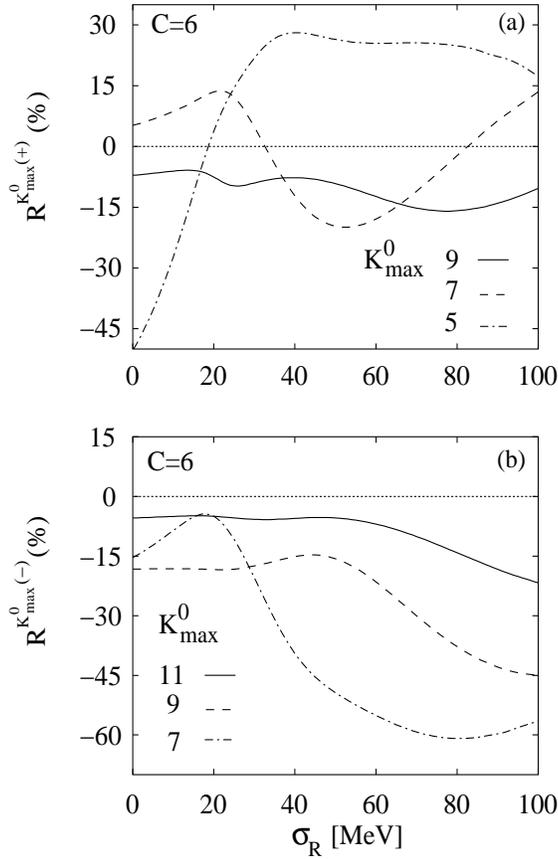}}}
\caption{Upper panel (a): Relative change $R^{K^0_{max}(+)}_C$ for the LIT of 
channel C=6 ($\sigma_I=10$ MeV). Lower panel (b): $R^{K^0_{max}(-)}_C$ for the 
same channel (see text for details).}
\label{fig1}
\end{figure}
As next point we address the quality of convergence of the LIT 
with respect to the HH expansion. For this reason, we first 
introduce the notation $\mathcal{L}^{K^0_{max}(\pm)}_C$ for the LIT calculated 
with an expansion up to $K^0_{max}$ for the ground state and up to 
$K^{\pm}_{max}=K^0_{max}\pm 1$ for $|\Psi_C\rangle$, respectively. 
That means $\mathcal{L}^{K^0_{max}(+)}_C$ represents the LIT of 
channel C calculated with a maximal hyperangular momentum for those 
states which can be reached by a dipole transition from the ground 
state, whereas for $\mathcal{L}^{K^0_{max}(-)}_C$ the expansion of 
channel C is taken up to one step below this value. Then we define 
\begin{equation}
R^{K^0_{max}(\pm)}_C=\frac{{\cal L}^{K^0_{max}(\pm)}_C-
{\cal L}^{(K^0_{max}-2)(+)}_C}
{{\cal L}^{K^0_{max}(\pm)}_C}\times 100,
\end{equation}
where $R^{K^0_{max}(+)}_C$ represents the relative percentage change 
for increasing $K^0_{max}$ by 2 together with a corresponding increase 
of $K_{max}$ while $R^{K^0_{max}(-)}_C$ represents the relative change
for increasing $K^0_{max}$ by 2 but keeping $K_{max}$ fixed. 

In Fig.~1~(a) we show the convergence pattern of $R^{K^0_{max}(+)}_C$
for one of the more important of the six $C$ channels, 
namely $C=6$. One readily sees that increasing $K^0_{max}$ results 
in a considerable reduction of $R^{K^0_{max}(+)}_6$. For example, changing 
$K^0_{max}$  from 7 to 9 yields a $R^{9(+)}_6$ between 5 and 15~\%, 
where the lower value refers to the region of the maximum of the LIT 
(compare Fig.~2), which 
however is still not negligible. It is very likely, that a further 
increase to $K^0_{max}=11$ would lead to a 
significant improvement of the convergence. Although the calculation of 
the ground state for $K^0_{max}=11$ can be performed, the corresponding
calculation of the final state with $K_{max}=12$ is beyond our present 
technical capabilities whereas it can be done for $K_{max}=10$, i.e.\
${\cal L}^{11(-)}_C$ can be calculated. 
Thus the question now is, how much is the convergence 
improved in going from ${\cal L}_C^{9(+)}$ to ${\cal L}_C^{11(-)}$. 
To this end we show $R^{K^0_{max}(-)}_C$ in Fig.~1~(b) for 
$K^0_{max}=7,\,9,\,11$. Again one notes a considerable decrease of 
$R^{K^0_{max}(-)}_C$ with increasing $K^0_{max}$ and for the highest
value $K^0_{max}=11$ of the order of 5~\% in the region $\sigma_R\leq 60$~MeV
where the LIT is more sizable. Furthermore, comparing panels (a) and (b)
of Fig.~1 one can see that for a given $K^0_{max}$  the relative change  
$R^{K^0_{max}(+)}_6$ is smaller than that of $R^{K^0_{max}(-)}_6$. Therefore, 
we expect that a further increase to  ${\cal L}_6^{11(+)}$  would lead to a  relative change with respect to ${\cal L}_6^{9(+)}$  smaller than $5\%$. 
Since the other channels behave similarly,
we estimate a total uncertainty of about $5\%$.

Now we turn to the results for photodisintegration. In order to obtain 
the total photoabsorption cross section $\sigma(\omega)$, we first  
invert the LITs of the response functions for the various channels
(for details see \cite{ELO99}). 
\begin{figure}[ht]
{\resizebox*{7cm}{15cm}{
\includegraphics{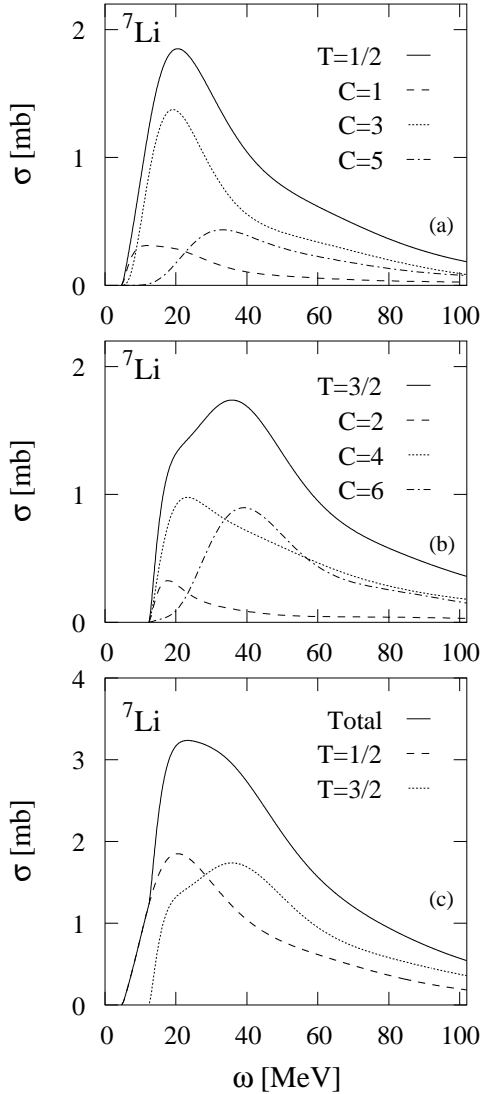}}}
\caption{Contribution of various channels to the total cross section.
Panels (a) and (b) show the separate contributions of the different 
channels and their sum for $T=1/2$ and $T=3/2$, respectively. Panel (c) 
shows again the $T=1/2$ and $T=3/2$ contributions and the total cross section.}
\label{fig2}
\end{figure}
\begin{figure}[ht]
{\resizebox*{9cm}{7cm}{
\includegraphics{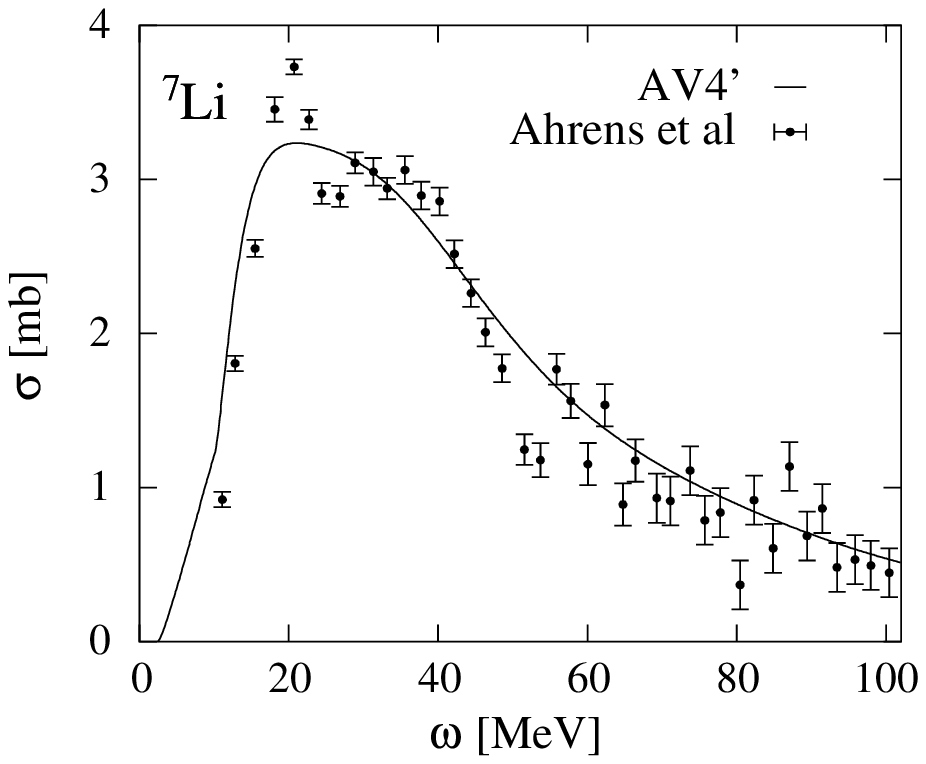}}}
\caption{Comparison of the theoretical photoabsorption cross section 
calculated with AV4' potential with experimental data from~\cite{Ahr75}.}
\label{fig3}
\end{figure}
Then one obtains the contribution to the cross section of each channel using
the relation (\ref{0}). Summing up all contributions finally yields the 
total cross section $\sigma(\omega)$. In Fig.~2 we show both the 
calculated total cross section and the contributions of the 
separate channels.
The energetically lowest open channel is the $T=\frac{1}{2}$ associated with
the reaction $^7$Li $+ \gamma \rightarrow$ $^4$He $+ t$ with a 
threshold of $2.47$ MeV (the theoretical threshold obtained with the AV4' 
potential is $4.70$ MeV), whereas the lowest open $T=\frac{3}{2}$ channel 
corresponds to $ ^7$Li $+ \gamma \rightarrow$ $^6$He $+$ p, whose 
threshold is 9.975 MeV (the theoretical value 
is 12.41 MeV). In Fig.~2 the theoretical thresholds have been used.

For the $T=1/2$-channels in Fig.~2~(a) one readily sees that by far 
the largest contribution comes from channel 3 rising steeply above 
threshold, reaching a maximum around 17~MeV and falling off only slowly. 
Channel 5 is the next in importance rising only slowly above 10~MeV with 
a maximum near 33~MeV and becoming then comparable in size to channel 3. 
Only in the very near threshold region channel 1 is dominant but then 
becomes much smaller than the other two channels. In Fig.~2~(b) the $T=3/2$ 
channels have two dominant contributions, almost similar in size. Channel 4
is slightly larger showing also a steep rise at threshold and slow fall-off
with a maximum near 23~MeV whereas the second in importance, channel 6, shows
a slow rise and a peak around 40~MeV. Compared to these two channels, 
the remaining channel 2 appears quite marginal. In view of the two maxima 
of almost equal height with a separation by about 17~MeV the total 
$T=3/2$-contribution exhibits a broader distribution than the 
$T=1/2$-contribution with a shoulder on the low-energy side. 
The maxima of both contributions have about the same size but are separated by 
about 20~MeV. Thus, the resulting total cross section in Fig.~2~(c) 
shows also a broad distribution with a steep rise right above threshold, 
a slight shoulder above the maximum and a slow fall-off at higher energies. 

This characteristic behavior is indeed exhibited by the experimental 
data on $^7$Li in Fig.~3 where we show a comparison of the theoretical 
result to experimental 
data from~\cite{Ahr75}. Note that the theoretical cross section is 
shifted here from the theoretical threshold to the experimental one. 
One readily notes that the gross properties of the data, steep rise, 
broad maximum and slow fall off, are very well reproduced quantitatively 
over the whole energy region by the theory. It is worthwhile to 
emphasize that this result is based on an ab initio calculation 
in which the complicated final state interaction of the 7N-system 
is rigorously taken into account by application of the LIT method. 
No adjustable parameters were used, the sole ingredient being the 
AV4' NN potential model. It remains to be seen whether the slight 
variation of the data near and above the maximum will also be found in 
an experiment with improved accuracy. Therefore, a new measurement of 
the total cross section with a higher precision would be very desirable. 
In particular, this could clarify the question whether a simple 
semi-realistic potential like the AV4' model is sufficient for 
an accurate theoretical description 
of this reaction or whether a more realistic nuclear force including a 
3N-force is needed.

\section*{Acknowledgments}  
A large part of the numerical calculations have been performed at the computer
centre CINECA (Bologna). This work was partially
supported by the Deutsche Forschungsgemeinschaft (SFB 443).
The work of N.B. was supported by the ISRAEL SCIENCE
FOUNDATION (grant no 202/02). 
We furthermore 
would like to thank J.\ Ahrens for providing us with the experimental data.

\end{document}